\documentclass[lettersize,journal]{IEEEtran}
\usepackage{amsmath,amsfonts}
\usepackage{algorithmic}
\usepackage{algorithm}
\usepackage{array}
\usepackage[caption=false,font=normalsize,labelfont=sf,textfont=sf]{subfig}

\usepackage{amsmath} 
\usepackage{amssymb}
\usepackage{amsfonts}
\usepackage{bm} 
\usepackage{fixmath} 
\usepackage{textcomp}
\usepackage{stfloats}
\usepackage{url}
\usepackage{verbatim}
\usepackage{graphicx}
\usepackage{cite}
\usepackage{bm}
\usepackage{enumitem}
\usepackage{cancel}
\usepackage{xcolor}

\newcommand{\h}{\boldsymbol{h}}

\newcommand{\Sb}{\boldsymbol{S}}
\newcommand{\A}{\boldsymbol{A}}
\usepackage{multirow}
\usepackage{tabularx}

\begin{document}

\title{ End-to-End Supervised Hierarchical Graph Clustering for Speaker Diarization}

\author{Prachi Singh,~\IEEEmembership{Student Member,~IEEE,} and  Sriram Ganapathy,~\IEEEmembership{Senior Member,~IEEE}
\thanks{This work is funded by grants from the British Telecom Research Center (BTIRC) and the Ministry of Information Technology (MEITY) Bhasini program. \\
The authors are with the Learning and Extraction of Acoustic Patterns (LEAP)
lab, Department of Electrical Engineering, Indian Institute of Science, Bangalore
560012, India.\\
E-mail: prachisingh@iisc.ac.in; sriramg@iisc.ac.in.}}


\maketitle

\begin{abstract}
Speaker diarization, the task of segmenting an audio recording based on speaker identity, constitutes an important speech pre-processing step for several downstream applications.  
The conventional approach to diarization involves multiple steps of embedding extraction and clustering, which are often optimized in an isolated fashion.
While end-to-end diarization systems attempt to learn a single model for the task, they are often cumbersome to train and require large supervised datasets. In this paper, we propose an end-to-end supervised hierarchical clustering algorithm based on graph neural networks (GNN), called End-to-end Supervised HierARchical  Clustering (E-SHARC). \textcolor{black}{The embedding extractor is initialized using a pre-trained x-vector model while the GNN model is trained initially using the x-vector embeddings from the pre-trained model. 
Finally, the E-SHARC model uses the  front-end mel-filterbank features as input and jointly optimizes the embedding extractor and the GNN clustering module,  performing representation learning, metric learning, and clustering with end-to-end optimization.}
Further, with additional inputs from an external overlap detector, the E-SHARC approach is capable of predicting the  speakers in the overlapping speech regions. 
The experimental evaluation on \textcolor{black}{benchmark datasets like AMI, Voxconverse and DISPLACE, 
illustrates that the proposed E-SHARC framework provides competitive diarization results using graph based clustering methods.}

\end{abstract}


\section{Introduction}
Speaker diarization is the process of automatically identifying and grouping audio segments based on speaker identity in the given recording. It is an important step in many speech-processing applications, such as meeting transcription, broadcast news analysis, speaker verification in conversational audio, 
call-center applications, etc. The key challenge in the task arises from short speaker turns, noise and reverberation, code-mixing/switching, and overlapping speech. 
In this paper, we propose a graph neural network-based representation learning and clustering framework for diarization of natural conversational audio.

\textcolor{black}{\textcolor{black}{Most of the diarization systems} segment the audio into short segments of $1-2$s, which are converted to embeddings, followed by a clustering step.
The initial efforts in embedding extraction framework used Gaussian mixture modeling based background models, termed as i-vectors \cite{dehak2010front}.}
Recent works in speaker diarization have focused on deep learning-based models such as Time-Delay Neural Networks (TDNN) \cite{snyder2018x,sell2018diarization} and \textcolor{black}{residual networks (ResNet)  \cite{He_2016_CVPR,zeinali2019but}}. These models can effectively extract speaker discriminative features, called x-vectors. The embeddings are used in the clustering step with a pre-defined similarity metric, e.g., Probabilistic Linear Discriminant Analysis (PLDA) scoring \cite{ioffe2006probabilistic,pldaivec}.
While this approach to diarization allows the use of pre-trained models (like embedding extraction networks and similarity scoring schemes) developed for speaker verification, the framework is highly modular, with each component optimized individually without a multi-speaker diarization loss function. 

Metric learning efforts attempt to partially mitigate this issue by learning the similarity metric based on diarization losses~\cite{Lin2019,narayanaswamy2019designing}. One of the limitations of these approaches is the difficulty in generalizing the learned similarity metric to new speakers and unseen conditions. This has motivated the work on self-supervised metric learning \cite{SSPIC,singh2021self}, which uses clustering outputs as the pseudo labels for model training. However, even with these efforts, the rest of the embedding extraction and clustering components are modular. 

The recent efforts in the development of an end-to-end neural diarization (EEND) system \cite{fujita2020neural,shinji2020Interspeech} \textcolor{black} {attempt to mitigate the issues in the modular systems}, 
where the goal is to perform the entire diarization process in a single neural network.  However, these models require large amounts of labeled conversational data for training and involve a large computational overhead. This can be difficult and time-consuming to obtain, and generalization to new unseen domains like multi-lingual or code-mixed settings may be challenging. 

To overcome the limitations of metric learning approaches, our prior work (Singh et al. \cite{singh2023supervised}) introduced a supervised hierarchical clustering approach called SHARC, which trains a  graph neural network with speaker embeddings as the nodes and similarity scores as the edges. End-to-end SHARC (E2E-SHARC)  was proposed to incorporate the joint learning of the embedding extractor and the graph neural network.
\textcolor{black}{This work was inspired by Xing et al. \cite{HGNN}, which performs supervised hierarchical clustering for images.} 

\textcolor{black}{This paper builds on the prior work in \cite{singh2023supervised} by providing thorough evaluation and analysis of the SHARC model. Further, this paper extends the E2E-SHARC model in a two-stage setting to address diarization of overlapping speech. The second stage is designed to identify the speaker identity of overlapping speaker. We refer to the overlap aware end-to-end SHARC model as E-SHARC-Overlap.}
\textcolor{black}{The framework allows the use of any deep learning-based 
embedding extractors, like Extended-TDNN (ETDNN) \cite{snyder2019speaker}, Factorized-TDNN (FTDNN) \cite{povey18_interspeech}, \textcolor{black}{ResNet \cite{zeinali2019but}} or ECAPA-TDNN \cite{dawalatabad2021ecapa} models}. 




The major contributions of the proposed work are:
\begin{itemize}
\item Provide a comprehensive mathematical and algorithmic description of graph-based clustering for speaker diarization task. 
\item \textcolor{black}{Provide detailed analysis of an end-to-end diarization system using supervised graph neural network-based clustering (E-SHARC).}
\item Introduce an overlap speaker detection approach \textcolor{black}{called E-SHARC-Overlap} to assign multiple speakers for the same audio region.
\item \textcolor{black}{Evaluate the performance on three benchmark datasets to illustrate improvements over state-of-the-art diarization systems with the oracle VAD (voice activity detection)}.
\end{itemize}

\section{Related work}
\subsection{\textcolor{black}{End-to-end neural diarization}}
\textcolor{black} {Horiguchi et al. \cite{shinji2020Interspeech} introduced a neural processing framework for speaker diarization.} The goal is to train a model that predicts the speaker activity labels at each time frame in an audio recording, given the front-end time-frequency features as input.  
\textcolor{black}{The model also attempts to identify the number of speakers in the conversation \textcolor{black}{using attractors based  on a long short-term memory} (LSTM)-based encoder-decoder network \cite{shinji2020Interspeech}.}
\textcolor{black}{To handle large number of speakers, the modeling in \cite{shinji2020Interspeech} was applied on short chunks and the speakers across chunks were clustered using a speaker similarity matrix updated using a loss in \cite{horiguchi2021towards}. Another approach called DiaPer \cite{landini2024diaper} replaces the encoder-decoder network with a \textcolor{black}{Perceiver-based decoder to incorporate  cross-attention for designing the attractor in a non-autoregressive manner. It is shown to perform better than the LSTM-based attractors for CALLHOME dataset \cite{callhome} in \cite{landini2024diaper}.}}
In contrast, our work does not require a large amount of supervised conversational training data. \textcolor{black} {Further, the E-SHARC framework is initialized with large pre-trained speaker embedding model which is further optimized for the diarization task.} 
\subsection{Graph clustering algorithms}
The task of grouping the vertices of the graph into clusters in such a way that there are more edges within each cluster and relatively few between the clusters is called \emph{graph clustering}. \textcolor{black}{Graph clustering has been largely unsupervised like spectral clustering \cite{spectral_clustering}.}  \textcolor{black}{The widely used clustering algorithm in speaker diarization is Agglomerative Hierarchical Clustering (AHC) \cite{AHC}. In AHC, the graph refers to a dendogram which illustrates the hierarchical merging process of clusters.}
Path integral clustering (PIC) \cite{PIC,SSPIC} is a graph-structural agglomerative clustering algorithm, where the graph encodes the structure of the feature space.   In our work, we describe the first efforts in exploring hierarchical graph clustering frameworks \cite{Lin2019,narayanaswamy2019designing} for speaker diarization.  We propose a supervised graph clustering framework that can be learned using the labeled data. 


\subsection{Overlap detection approaches}
Recent diarization systems have relied on external overlap detection methods to perform overlap speaker assignments. Bullock et al. \cite{overlapbullock} and Sajjan et al. \cite{sajjan2018leveraging} proposed the Long Short-Term Memory (LSTM) based architecture for overlap detection. 
\textcolor{black}{The model in Pyannote toolkit~\cite{Bredin2020,bredin21_interspeech} is trained with hand-crafted features or trainable SincNet features. The model performs speaker activity detection for $5$s chunks and then extracts homogeneous speaker segment embeddings for each chunk, which are used to cluster the  segments to perform diarization. The speech activity of multiple speakers at the same time indicates overlapped speech.
In contrast, our work performs graph clustering of the uniform speech segments embeddings in the entire audio during the first pass. The second pass utilizes overlapped speech regions detected by Pyannote toolkit to determine the second speaker's identity.} 

\section{Proposed Approach}
\textcolor{black} {In this section, we describe the proposed E-SHARC model. It involves two steps, namely embedding extraction and GNN-based clustering. In this work, we have explored the ETDNN model based embedding extractor. The GNN-based clustering (Section \ref{sec:sharc_clustering}) and GNN module training (Section \ref{sec:model_training}.2) are motivated by the work of  Xing et al. \cite{HGNN}.}
\subsection{Background}
Graphs are powerful mathematical structures that represent complex relationships and interactions among entities. Graph algorithms find applications in various areas because of their ability to exploit the structure of data, which can be in any form \cite{bondy1991graph, majeed2020graph}. 
Graph frameworks are suitable for diarization problems due to their ability to represent speaker interactions in a conversation naturally.

 A graph $\pmb{\mathbb{G}}$ can be described by the set of vertices/nodes $\pmb{\mathbb{V}}$ and edges $\pmb{\mathbb{E}}$, as $\pmb{\mathbb{G}}=(\pmb{\mathbb{V}}, \pmb{\mathbb{E}})$.
The adjacency matrix \textcolor{black}{($\A \in \mathbb{R}^{N \times N}$, $N$ - number of nodes)} captures connections between nodes. 
The adjacency matrix can be obtained using similarity scores between data points. \textcolor{black}{In the context of diarization, nodes in the graph represent embeddings of individual speakers, and edges capture the turn-taking behavior among the   speakers in the conversation.} 

A graph neural network (GNN) is an effective tool for analyzing and modeling graphs. 
\textcolor{black}{The Graph Convolution Network (GCN) \cite{kipf2016semi}, the most common variant of GNNs, was used in \cite{TongGCN} for semi-supervised training. A GNN was used to cluster embeddings corresponding to unlabeled data and the ``pseudo-labels" were used to retrain the embedding extractor.} Speaker diarization can be formulated as a link prediction problem \cite{GNNfordiarization} between speaker embeddings of segments from the same recording. The GCN is inherently transductive and does not generalize to unseen nodes. The GraphSAGE \cite{hamilton2017inductive}, another variant of GNN, is a representation learning technique suitable for dynamic graphs. Without the need for retraining, it can predict the embedding of a new node. Therefore, we have explored this variant in the current work.
 

\subsection{Notations} 
\textcolor{black} {Following is the list of symbols used across the paper to describe the input, output and parameters of the model. Some of the symbols will be explained in later sections based on the context.}
\begin{itemize}[leftmargin=*]
    \item $\pmb{\mathbb{{X}}}_r = \{\bm{x}_{1},...,\bm{x}_{N_r}\}, \bm{x}_n \in \mathbb{R}^{D} ~\forall n\in \{1,...,N_r\}$ denotes the sequence of $N_r$  segment embeddings for the recording $r$.
    \item $m\in\{1,2,...,M\}$ denotes the level (iteration) of hierarchical clustering.
      \item \textcolor{black}{$\pmb{\mathbb{G}}^{(m)}_r=(\pmb{\mathbb{V}}^{(m)}_r, \pmb{\mathbb{E}}^{(m)}_r)$ is the graph of recording $r$ at level $m$. $\pmb{\mathbb{V}}^{(m)}_r$ and $\pmb{\mathbb{E}}^{(m)}_r$ are the set of vertices and set of edges of the graph, respectively.}
      
      \item  $\h_i^{(m)}= [{\Tilde{\h}_i}^{(m)};{\Bar{\Bar{\h}}_i}^{(m)}]$ where ${\Tilde{\h}_i}^{(m)}$ is the identity feature of node $i$ \textcolor{black}{at level $m$. 
      It is the representative feature of the node with the highest density among all nodes of the previous level that were clustered to form $i$.}
      
      ${\Bar{\Bar{\h}}_i}^{(m)}$ is the average feature of level $m$, which is the average of all the identity features from the previous level $m-1$ (Equation \ref{eqn:clusterfeat}).  $\h_i^{(0)} = [\bm{x}_{i};\bm{x}_{i}]$.
     
      \item $\pmb{\mathbb{H}}^{(m)}_r = \{\h_1^{(m)}, \h_2^{(m)},...,\h_{N^{(m)}_r}^{(m)}\} \in \mathbb{R}^{D'\times N^{(m)}_r}$ denotes the node features. \textcolor{black}{Here, $D'=2D$ is the dimension of node features}, $N^{(m)}_r$ is the number of nodes at level $m$ for recording $r$ and $N^0_r=N_r$. 
       \item $\Sb^{(m)}_r\in\mathbb{R}^{N^{(m)}_r \times N^{(m)}_r}$ denotes pairwise similarity score matrix such that $[\Sb^{(m)}_r]_{ij}=s({\Tilde{\h}_i}^{(m)},{\Tilde{\h}_j}^{(m)})$, the similarity score between identity features of nodes $i$ and $j$ at level $m$. 

   \item $p^{(m)}_{ij}$ and $\hat{p}^{(m)}_{ij}$ denote the ground truth and predicted edge probabilities between node $i$ and $j$, respectively, at level $m$. ${P}^{(m)}$ and $\hat{P}^{(m)}$ are the ground truth and predicted edge sets, respectively, as shown in Figure \ref{fig:trainschematic}.
   
   \item $\hat{e}^{(m)}_{ij}=2\hat{p}^{(m)}_{ij}-1 \in [-1,1]$ denotes the edge coefficient or edge weight between node $i\in\{1,..,N^{(m)}_r\}$ and $j\in J^k_i$, where $J^k_i$ represents the set of $k$-nn ($k$-nearest neighbor) nodes of $i$ at level $m$. 
    \item \textcolor{black}{$\pmb{\mathbb{C}}_r^{(m)}=\{\boldsymbol{\mathcal{C}}_1^{(m)},...,\boldsymbol{\mathcal{C}}_{n_c}^{(m)}\}$} denotes the set of clusters for level $m$ and recording $r$ where, each cluster set $\boldsymbol{\mathcal{C}}_i^{(m)}$ contains all connected components/nodes without any discontinuity, $n_c$ is the total number of clusters formed. $N_r^{(m+1)}=n_c$.
    \item Hyper-parameters: \\
   $k$ - number of nearest-neighbors, \\
  $\tau$ - threshold on edge probabilities to stop merging.\\
\textcolor{black}{ The symbols for a given recording are denoted with subscript $r$ during training  or $t$ at test time.  However, in some expressions  below, we have not notated it explicitly.}

\end{itemize}





\begin{figure*}[t!]
  \centering
  \includegraphics[trim={0cm 0cm 1cm 0cm},clip,width=\textwidth]{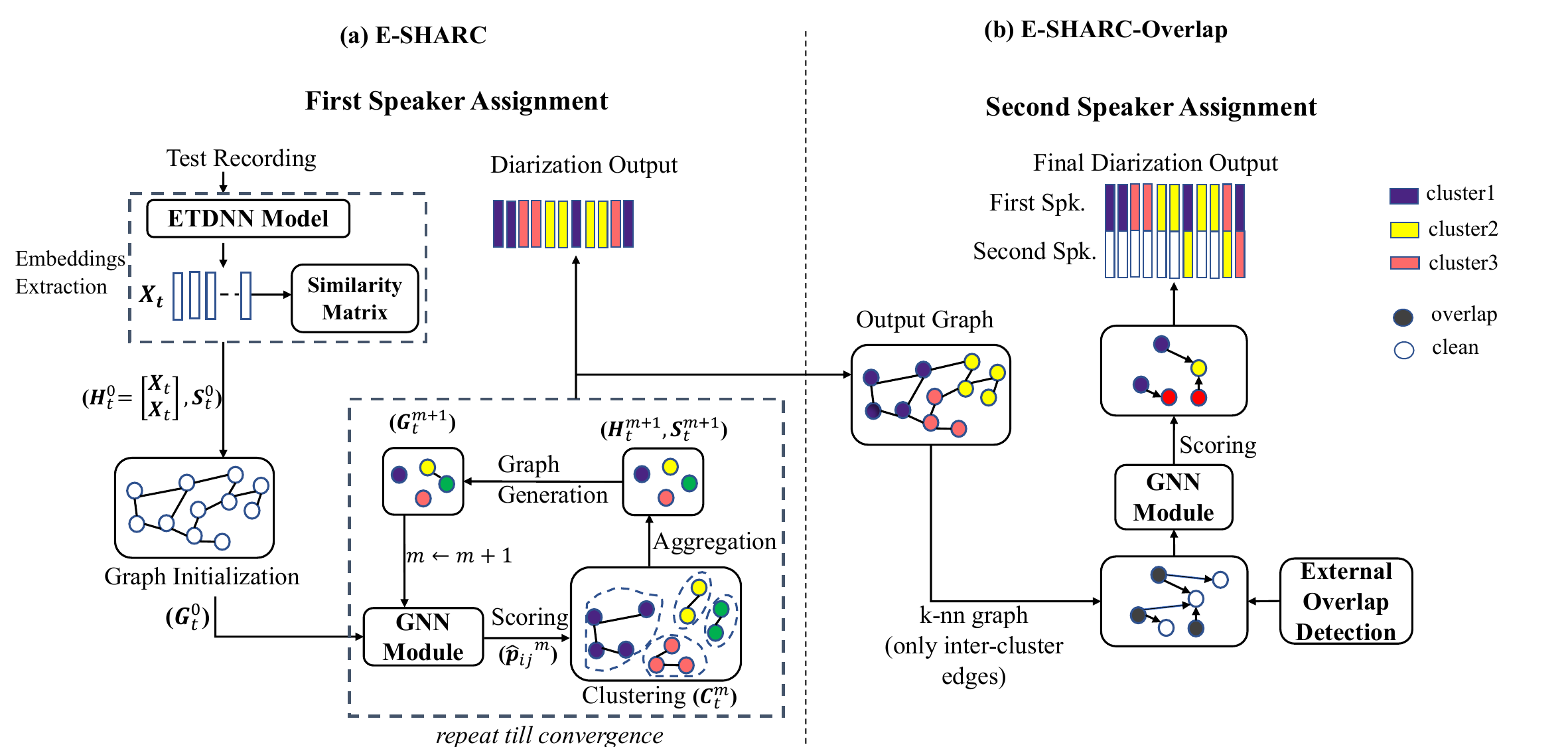}
  \caption{Block schematic of the E-SHARC algorithm with overlap handling. (a) shows E-SHARC inference containing ETDNN and GNN modules for the first speaker assignment. (b) shows E-SHARC-Overlap for the second speaker assignment approach using an external overlap detector and the GNN module. \textcolor{black}{The arrows points to the k-nearest neighbors of a node after removing the intra-cluster edges. }}
  \label{fig:schematic}
\end{figure*}

\subsection{Graph initialization}
For both the training and inference steps of the E-SHARC framework, the first step is the creation of a graph based on the input embeddings (x-vectors). The initial graph, termed as the graph at level $m=0$, $\pmb{\mathbb{G}}_t^{(0)}=(\pmb{\mathbb{V}}_t^{(0)},\pmb{\mathbb{E}}_t^{(0)})$, contains embeddings as the nodes in $\pmb{\mathbb{V}}_t^{(0)}$  and $k$-nearest neighbor of each node based on $\Sb_t^{(0)}$ similarity score matrix that forms the edges in  $\pmb{\mathbb{E}}_t^{(0)}$. A pre-trained PLDA model is used to generate similarity scores.

\subsection{Forward pass}
\label{sec:sharc_clustering}

Figure \ref{fig:schematic} (a) shows the block diagram of the inference step. For a test recording $t$,  x-vectors $\pmb{\mathbb{X}}_t$ are extracted to perform hierarchical clustering.  The SHARC is performed using a GNN module, which takes the graph as input and generates edge prediction probabilities. 

The components of the SHARC algorithm are discussed below.
\subsubsection{GNN scoring }
\label{sec:gnnscoring}
The GNN scoring function $\Phi$ is a learnable GNN module designed for supervised clustering. The module jointly predicts node densities and edge probabilities using the input embeddings at each level. Each graph $\pmb{\mathbb{G}}^{(m)}_t$, 
 containing source and destination node pairs, is fed to the GNN scoring model. 
 The output of the model is linkage probability $\hat{p}^{(m)}_{ij}$ of the nodes $v^{(m)}_i$ and $v^{(m)}_j ~\forall i={1,..,N^{(m)}_t}, j\in J^k_i$. 


The edge probabilities are also used to compute node density, which measures how densely the node is connected within the cluster. A node with higher density better represents the cluster than a node with lower density. 
A node density using predicted edge coefficients is called pseudo density $\hat{d}^{(m)}_i$ for node $v^{(m)}_i$ and is defined as:
\vspace{-6pt}
\begin{equation}
    \hat{d}^{(m)}_i = \frac{1}{k} \sum^k_{j\in J^k_i}{\hat{e}^{(m)}_{ij}\Sb^{(m)}_t(i,j)}
\end{equation}
The ground truth density $d^{(m)}_i$ is obtained using ground truth edge coefficient $e^{(m)}_{ij}= 2p^{(m)}_{ij} - 1\in \{-1,1\}$,  
where $p^{(m)}_{ij}=1$ if nodes $v^{(m)}_i$ and $v^{(m)}_j$ belong to the same cluster/speaker, otherwise $p^{(m)}_{ij}=0$. 

\subsubsection{Clustering}
\label{sec:clustering}
Clustering is the process of grouping the nodes based on the presence of edge connections. After GNN scoring, clustering is performed hierarchically 
using the edge probabilities  $\hat{p}^{(m)}_{ij}$ and the estimated node densities $\hat{d}_i^{(m)}$. 
At each level of hierarchy $m$, it creates a candidate 
edge set $\varepsilon (i)^{(m)}$, for the node $v^{(m)}_i$, with edge connection threshold $\tau$ as,
\begin{equation}
\label{eqn:edge_set}
\begin{split}
\varepsilon (i)^{(m)} =&\{j|(v^{(m)}_i,v^{(m)}_j) \in E_t^{(m)},\\
&\quad \hat{d}^{(m)}_i \leq \hat{d}^{(m)}_j \quad \textrm{and}\quad \hat{p}^{(m)}_{ij} \geq \tau \} 
\end{split}
\end{equation}
For any $i$, if $\varepsilon(i)^{(m)} \neq \varnothing$ , pick $j = $ \textit{argmax}$_{j \in \varepsilon (i)^{(m)}} ~\hat{e}^{(m)}_{ij}$ and connect $v^{(m)}_i$ and $v^{(m)}_j$. After a full pass over every node, a set of clusters $\pmb{\mathbb{C}}_t^{(m)}$ is formed based on connected components.

\subsubsection{Feature aggregation}
\label{sec:feat_aggre}
The cluster nodes at level $m$ are aggregated to form node features of the next level.  
To obtain node representations for next level $\pmb{\mathbb{H}}_t^{(m+1)}$, the clusters $\pmb{\mathbb{C}}_t^{(m)}$ and the features $\pmb{\mathbb{H}}_t^{(m)}$ are used to compute identity feature ${\Tilde{\h}_i}^{(m+1)}$ and average feature ${\Bar{\Bar{\h}}_i}^{(m+1)}$ of each cluster $i$ as, 
\begin{equation}
\label{eqn:clusterfeat}
\begin{split}
{\Tilde{\h}_i}^{(m+1)} &= {\Tilde{\h}^{(m)}}_{z_i} ; \quad \quad
{\Bar{\Bar{\h}}_i}^{(m+1)} = \frac{1}{|{\boldsymbol{\mathcal{C}}_i}^{(m)}|}\sum_{j \in {\boldsymbol{\mathcal{C}}_i}^{(m)}} {\Tilde{\h}_j}^{(m)} \\
{\h_i}^{(m+1)} &= [{\Tilde{\h}_i}^{(m+1)};{\Bar{\Bar{\h}}_i}^{(m+1)}]
\end{split}
\end{equation}
where $z_i$ = \textit{argmax}$_{j \in \mathcal{C}_i^{(m)}} ~{\hat{d}_j}^{(m)}$.

\subsubsection{Graph generation}
A new graph $\pmb{\mathbb{G}}_t^{(m+1)}$ is constructed for the next level using the node features $\pmb{\mathbb{H}}_t^{(m+1)}$. The edges are formed using $\Sb_t^{(m+1)}$ which is computed using the identity features $\pmb{\mathbb{H}}_t^{(m+1)}$.

The algorithm repeats until convergence when there are no connected components in the graph, as shown in Figure \ref{fig:schematic} (a).

\begin{figure}[t!]
  \centering
  \includegraphics[trim={0cm 0cm 17cm 0cm},clip,width=\columnwidth]{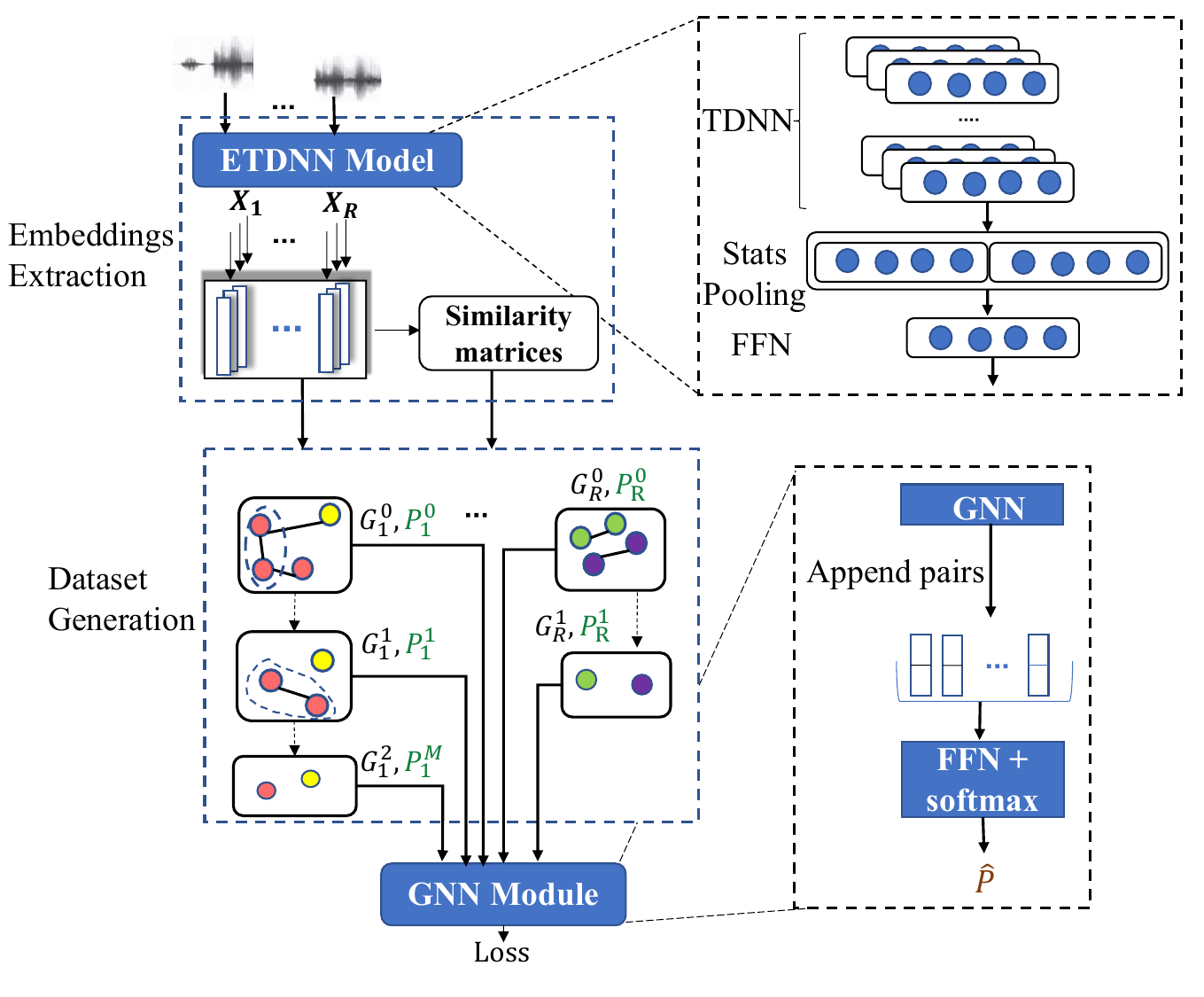}
  \vspace{-0.7cm}
  \caption{Block schematic of the E-SHARC training. The ETDNN and GNN modules in black blocks contain learnable parameters. The GNN module generates edge probabilities and weights used in loss computation.}
    \vspace{-10pt}
  \label{fig:trainschematic}
\end{figure}

\subsection{Model training}
\label{sec:model_training}

\textcolor{black}{The stages involved in E-SHARC model training are},
\begin{enumerate}
    \item \textcolor{black}{\textbf{Pre-training} -  Train the x-vector module using speaker classification loss. Train PLDA using the x-vector embeddings and the speaker labels.}
    \item \textcolor{black}{ \textbf{Data preparation} - Preparing the training data in the form of graph inputs and outputs for GNN module training using x-vectors as node vectors and edge connections defined using PLDA.}
    \item \textcolor{black}{ \textbf{GNN module training} - Train the GNN module using the SHARC training procedure.}
    \item \textcolor{black}{ \textbf{E-SHARC training} - Train the joint model of x-vector embedding network and the GNN using the graph clustering loss.  }
\end{enumerate}


\textcolor{black}{The stage 1 is same as our baseline system which is described in Section \ref{sec:baseline} and the stages 2-4 are described in detail below.} \\
\textbf{Data preparation} - Multiple graphs are constructed using input features and similarity matrices at different levels of clustering for each of the audio recordings in the training set. The loss combines binary cross entropy loss and mean squared error loss.
These losses across all graphs are then accumulated and back-propagated for each batch. The block diagram is shown in Figure \ref{fig:trainschematic}.

For the training set, input graphs $\pmb{\mathbb{G}}= \{\pmb{\mathbb{G}}^{(0)}_{1}, \pmb{\mathbb{G}}^{(1)}_{1}, ..., \pmb{\mathbb{G}}^{(1)}_{1}, ..., \pmb{\mathbb{G}}^{(M_r)}_{R}\}$ are constructed at different clustering levels for each recording $r$.  $M_r$ is the maximum number of levels created for $r$.  $\pmb{\mathbb{E}}=\{\pmb{\mathbb{E}}_1^{(0)},..,\pmb{\mathbb{E}}_1^{(M_1)}, ..., \pmb{\mathbb{E}}_R^{(0)},..,\pmb{\mathbb{E}}_R^{(M_R)}\}$ and $\pmb{\mathbb{V}}=\{\pmb{\mathbb{V}}_1^{(0)},..,\pmb{\mathbb{V}}_1^{(M_1)}, ..., \pmb{\mathbb{V}}_R^{(0)},..,\pmb{\mathbb{V}}_R^{(M_R)}\}$ are the set of all possible edges and nodes, respectively. 
\textcolor{black}{For each level $m$, graph $\pmb{\mathbb{G}}^{(m)}_{r}$ consists of nodes that belong to the same speaker cluster and satisfies the condition in Equation \ref{eqn:edge_set} with $\tau=0$. The node features are obtained using Equation \ref{eqn:clusterfeat}.}


\noindent  \textcolor{black}{\textbf{GNN module training}} - 
This step uses a pre-trained extended time delay neural network (ETDNN) \cite{snyder2019speaker} model for embedding extraction to extract x-vectors. These x-vectors are used to generate the training graphs. In the first step, the weights of the ETDNN module are frozen, and the GNN module weights are learned.

The GNN model consists of one GNN layer with $D''=2048$ units (neurons in a layer).  
It takes  node representations $\pmb{\mathbb{H}}_r^{(m)}$
 and their edge connections $\pmb{\mathbb{E}}^{(m)}_r$ as input and generates latent representations denoted as \textcolor{black}{$\pmb{\mathbb{\hat{H}}}_r^{(m)} \in \mathbb{R}^{D''\times N_r^{(m)}}$}. Each node $v_i$ associates with a cluster (speaker) label $z_i$ in the training set, allowing the learning of the clustering criterion from the data. The pair of embeddings are concatenated $[\hat{\h}_i;\hat{\h}_j]$ and passed to a fully connected feed-forward network with a size of $\{2D'',1024,1024,2\}$ followed by softmax activation to generate edge probability $\hat{p}_{ij}$.  The architecture was selected based on validation experiments.

 The GNN module is trained using the following loss function.
\vspace{-6pt}
\begin{equation}
\label{eqn:tot_loss}
L = L_{conn} + L_{den}
\end{equation}
where $L_{conn}$ is the pairwise binary cross entropy loss based on edge probabilities across all the possible edges in  $\pmb{\mathbb{E}}$ given as (\textcolor{black}{as described in \cite{HGNN}}) :
\begin{equation}
\label{eqn:link_loss}
L_{conn} = -\frac{1}{|\pmb{\mathbb{E}}|} \sum_{(v_i,v_j) \in \pmb{\mathbb{E}}} l_{ij}
\end{equation}
\begin{equation}
\label{eqn:edge_loss}
l_{ij} =
\left \{
    \begin{array}{ll}
		p_{ij}\log \hat{p}_{ij} + (1 - p_{ij})\log (1 - \hat{p}_{ij})  & \mbox{if } d_i \leq d_j \\
		0 & \mbox{otherwise }
	\end{array}
\right.
\end{equation}
Here, $|\pmb{\mathbb{E}}|$ represents the total number of edges.
$L_{den}$ represents the neighborhood density average loss given by Equation \ref{eqn:den_loss}. 
$L_{den}$ represents mean squared error  (MSE)  loss between ground truth node density $d_i$ and predicted density $\hat{d}_i$,
\begin{equation}
\label{eqn:den_loss}
L_{den} = \frac{1}{|\pmb{\mathbb{V}}|} \sum_{i=1}^{|\pmb{\mathbb{V}}|} ||d_i - \hat{d}_i||_2^2
\end{equation}
where $|\pmb{\mathbb{V}}|$ is the cardinality of $\pmb{\mathbb{V}}$.\\ 
\textbf {E-SHARC training} - 
\label{ETDNNarchitecture} \textcolor{black}{All the model parameters including
 the $13$-layer ETDNN model, and the GNN model are fine-tuned using the final loss (Eq.~\ref{eqn:tot_loss}).
}
\textcolor{black}{Following the E-SHARC training, we do not retrain the PLDA model as this was not found to improve the results further. The continuous update of the  PLDA model parameters along with the SHARC training may potentially improve the modeling. However, in this paper, we have not updated the PLDA model.}
\section{Handling Overlapped Speech}
We extend the E-SHARC model also to perform GNN-based overlap prediction called E-SHARC-Overlap. Our approach assumes that a maximum of two speakers are present in an overlapping region.
We follow a two-pass approach to accurately identify the speakers in the overlapping region. 
\textcolor{black}{During training, the first pass comprises E-SHARC modeling on the clean segments (single speaker segments).} The first speaker, referred to as the parent cluster, is selected for each segment based on the first-pass E-SHARC algorithm. The proposed overlap model is trained using overlapping and clean segments in the second pass.
 The graph adjacency matrix is generated by connecting the nodes from one cluster to nodes from any other cluster except the parent cluster. This enables the overlap model to identify the second speaker in the overlapping regions. A small percentage of intra-cluster connections ($10\%$)  are preserved randomly, enabling contrastive training.
The k-nearest neighbors of each node are selected based on the final graph adjacency matrix. The training loss comprises BCE and MSE losses, similar to the original SHARC model (Equation \ref{eqn:tot_loss}).

\textit{Inference}\\
The inference steps comprise first pass - clustering, second pass - overlap detection, and second speaker assignment to generate the final diarization output (Figure \ref{fig:schematic}). 
\subsubsection{First speaker assignment}
First-pass clustering based on E-SHARC is performed to assign an initial parent cluster and to decide the number of speakers present in a recording, as described in Section \ref{sec:sharc_clustering}.  The input x-vectors are the average of the pre-trained x-vectors and the E-SHARC x-vectors.

\subsubsection{Second/Overlap speaker assignment}
The overlap or second speaker assignment is done using the E-SHARC-Overlap model. The pyannote overlap detector \cite{Bredin2020, overlapbullock} is used to identify regions containing overlapped speech. 
A new graph is created using overlapping nodes and their $k$-nn after removing the within-cluster connections.
Then, for each node, top $k'$ ($k' \le k$) neighbors are selected based on edge probabilities from the E-SHARC-Overlap model. The dominant cluster identity of the neighbors is assigned as the second speaker for the node.


\section{Experiments}
\subsection{Datasets}
\label{sec:datasets}
\begin{itemize}[leftmargin=*]
	\item \textbf{The AMI dataset}: 
	The ofﬁcial speech recognition partition of the AMI dataset ~\cite{mccowan2005ami} comprises training, development (dev), and evaluation (eval) sets consisting of $136$, $18$, and $16$ recordings sampled at $16$kHz, respectively.
	The single-distant microphone (SDM) condition of the AMI dataset is used for experiments. It contains single-channel microphone recordings. \textcolor{black}{The AMI train set contains 75 hrs of labeled speech.
	The number of speakers and the duration of each recording is in the range 3-5 and $20$-$60$ mins, respectively.}

	\item \textbf{The Voxconverse dataset}: It is an audio-visual diarization dataset\cite{chung20_interspeech} \textcolor{black}{consisting of multispeaker speech} recordings extracted from YouTube videos. It is divided into a development (dev) set and an evaluation (eval) set consisting of $216$ and $232$ recordings, respectively. The duration of a recording is in the range $22$-$1200$ s. The number of speakers per recording ranges from $1$ to $21$. \textcolor{black}{The dataset used for training is simulated using Voxceleb 1 and Voxceleb 2 \cite{nagrani2017voxceleb,Chung2018} and Librispeech \cite{panayotov2015librispeech} following the recipe mentioned in \cite{shinji2019ASRU}.
We simulated $5000$ mixtures containing $2$-$5$ speakers with duration ranging from $150$ to $440$ s. This generates about $400$ hrs of data with $6,023$ speakers. }
\item{\textbf{The DISPLACE challenge dataset:}}
The DISPLACE 2023 \cite{baghel2023displace} dataset released as part of the challenge comprises single-channel far-field natural multilingual, multi-speaker conversational speech recordings. The development (dev) and evaluation phase 2 (eval) sets contain 27 recordings (15.5 hours) and 29 recordings (16 hours), respectively. The speakers in the dev and eval sets are mutually exclusive. The dataset contains conversations in Hindi, Kannada, Bengali, Malayalam, Telugu, Tamil, and Indian English. In each recording,  the number of speakers ranges from $3$ to $5$, and the number of languages ranges from $1$ to $3$.

\end{itemize}

\subsection{Baseline system}
\vspace{-2pt}
\label{sec:baseline}
The baseline method is an x-vector clustering-based approach described in \cite{ryant21_interspeech,singh2021self}.
First, the recording is divided into 1.5s segments with 0.75s shift. The 40D mel-spectrogram features are computed from each segment which is passed to the ETDNN model \cite{snyder2019speaker} to extract 
512D x-vectors. The ETDNN model is trained on the VoxCeleb1 \cite{nagrani2017voxceleb} and VoxCeleb2 \cite{Chung2018} datasets, for the speaker identification task, to discriminate among the $7,146$ speakers.
 The input to the model is 40-D mel-spectrogram features extracted from each segment (1.5s) of the training recording followed by cepstral mean normalization. The model comprises four blocks containing  TDNN and a fully connected layer of size 1024D. This is followed by two feed-forward layers containing $\{1024, 2000\}$ units and a segment pooling layer of 4000D concatenating the mean and standard deviation of each segment from the previous layer. This is passed to an affine layer of 512D which is called the x-vector embedding. 
The whitening transform, length normalization, and recording level PCA (dimension=30) are applied to the x-vectors as pre-processing steps. 
These embeddings are used to compute the PLDA similarity score matrix and clustered to generate speaker labels for each segment. The PLDA model is trained using the x-vectors. 

For the baseline system, we have used the two most popular clustering approaches - Agglomerative Hierarchical Clustering (AHC) \cite{AHC} and Spectral Clustering (SC) \cite{spectral_clustering}. The PLDA similarity scores are used directly to perform AHC. For SC, we convert the PLDA scores $s$ to $s' \in [0,1]$ by applying sigmoid with temperature parameter $\alpha=0.1$ (best value obtained from experimentation) as: $s'=\frac{1}{1+exp(-s/\alpha)}$. This helps to improve the performance without other pre-processing steps usually followed in previous implementations \cite{Lin2019}.

\begin{table}[t]
\caption{Choice of hyper-parameters (HP) for train, dev, eval split of AMI and Voxconverse datasets.}
\label{tab:hyperparameters}
\centering
\resizebox{\columnwidth}{!}{
\begin{tabular}{|l|l|lll|lll|}
\hline
\multirow{2}{*}{\textbf{Model}} & \multirow{2}{*}{\textbf{HP}} & \multicolumn{3}{c|}{\textbf{AMI}}                                     & \multicolumn{3}{c|}{\textbf{Voxconverse}}                              \\ \cline{3-8} 
                           & & \multicolumn{1}{l|}{Train} & \multicolumn{1}{l|}{Dev} & Eval & \multicolumn{1}{l|}{Train} & \multicolumn{1}{l|}{Dev} & Eval \\ \hline
 SHARC & $k$                           & \multicolumn{1}{l|}{60}    & \multicolumn{1}{l|}{60}  & 60   & \multicolumn{1}{l|}{60}    & \multicolumn{1}{l|}{30}  &30   \\
 SHARC & $\tau$                & \multicolumn{1}{l|}{-}     & \multicolumn{1}{l|}{0.0} & 0.0  & \multicolumn{1}{l|}{-}     & \multicolumn{1}{l|}{0.5} &0.8  \\
E-SHARC & $k$                     & \multicolumn{1}{l|}{30}    & \multicolumn{1}{l|}{50}  & 50   & \multicolumn{1}{l|}{60}    & \multicolumn{1}{l|}{30}  & 30   \\
E-SHARC & $\tau$                  & \multicolumn{1}{l|}{-}     & \multicolumn{1}{l|}{0.0} & 0.0  & \multicolumn{1}{l|}{-}     & \multicolumn{1}{l|}{0.9} & 0.8  \\ \hline
\end{tabular}
}
\end{table}

\begin{table}[t]
\caption{Performance comparison based on cluster purity and coverage for Voxconverse dataset.}
\label{tab:purityCov}
\centering
\resizebox{\columnwidth}{!}{
\begin{tabular}{|l|c|c|}
\hline
\textbf{Method}   & \textbf{Cluster Purity} & \textbf{Cluster Coverage} \\ \hline
Baseline with AHC & \textbf{93.5}            & 89.5              \\
Baseline with SC  & 92.0            & 92.3              \\
SHARC             & 93.0            & 92.4              \\
E-SHARC           & 93.0            & \textbf{92.9}             \\ \hline
\end{tabular}}
\vspace{-0.1in}
\end{table}

 
 \begin{table*}[t]
 \caption{DER (\%) comparison on the AMI, Voxconverse, and DISPLACE datasets with the baseline methods considering overlaps and without tolerance collar.}
 \label{tab:ovpresults}
 \resizebox{\textwidth}{!}{
 \begin{tabular}{|l|llll|llll|llll|}
 \hline
 \multirow{2}{*}{\textbf{System}} &
   \multicolumn{4}{c|}{\textbf{AMI Eval}} &
   \multicolumn{4}{c|}{\textbf{Voxconverse Eval}} &
   \multicolumn{4}{c|}{\textbf{DISPLACE Eval}} \\ \cline{2-13} 
  &
   \textbf{FA} &
   \textbf{Miss} &
   \multicolumn{1}{l|}{\textbf{Conf.}} &
   \textbf{DER} &
   \textbf{FA} &
   \textbf{Miss} &
   \multicolumn{1}{l|}{\textbf{Conf.}} &
   \textbf{DER} &
   \textbf{FA} &
   \textbf{Miss} &
   \multicolumn{1}{l|}{\textbf{Conf.}} &
   \textbf{DER} \\ \hline
 AHC             & 0.0  & 15.6   &  \multicolumn{1}{l|}{13.9} & 29.51 & 0.0 & 3.1 & \multicolumn{1}{l|}{10.31} & 13.41 & 3.1 & 22.4 & \multicolumn{1}{l|}{15.0} & 40.60 \\
 AHC+Overlap     & 0.7 & 11.6 &  \multicolumn{1}{l|}{14.37} & 26.67 & 0.8  & 1.6 & \multicolumn{1}{l|}{9.65} & 12.05 & 3.8 & 21.0  & \multicolumn{1}{l|}{15.67} & 40.47      \\ \hline
 SC              & 0.0 & 15.6 &  \multicolumn{1}{l|}{6.7} & 22.29 & 0.0 & 3.1 & \multicolumn{1}{l|}{10.9} & 14.02 & 3.1 & 22.4 & \multicolumn{1}{l|}{15.3} & 40.84 \\
 SC+Overlap   & 0.7  & 11.6 & \multicolumn{1}{l|}{8.06 } & 20.36  & 0.8  & 1.6 & \multicolumn{1}{l|}{11.33} & 13.73   & 3.8 & 21.0 & \multicolumn{1}{l|}{15.85} & 40.65      \\ \hline
 SHARC           & 0.0 & 15.6 &  \multicolumn{1}{l|}{5.7 } & 21.27 & 0.0 & 3.1 & \multicolumn{1}{l|}{10.2} & 13.29 & 3.1 & 22.4 & \multicolumn{1}{l|}{7.47} & 33.07 \\
 \textcolor{black}{+ SHARC-Overlap}   & 0.7  & 11.6 & \multicolumn{1}{l|}{7.2} & 19.50  & 0.8 & 1.6 & \multicolumn{1}{l|}{10.16} & 12.56 & 3.8 & 21.0 & \multicolumn{1}{l|}{ 7.9 } & 32.73   \\ \hline
 E-SHARC         & 0.0 & 15.6  & \multicolumn{1}{l|}{4.3} & 19.83 & 0.0 & 3.1  & \multicolumn{1}{l|}{8.5} & 11.68 & 3.1 & 22.4 & \multicolumn {1}{l|}{7.33} & 32.93 \\
 
 \textcolor{black}{+ E-SHARC-Overlap}       & 0.7 & 11.6  & \multicolumn{1}{l|}{5.69} & \textbf{17.99}  & 0.8 & 1.6  & \multicolumn{1}{l|}{9.02} & \textbf{11.42} &  3.8 & 21.0 & \multicolumn{1}{l|}{ 7.65} &\textbf{32.45}  \\ \hline
 \end{tabular}}
 \end{table*}

\begin{table*}[t]
 \caption{DER (\%) comparison on the AMI, Voxconverse, and DISPLACE datasets with the baseline methods after VBx resegmentation. Considering overlaps and without tolerance collar.}
 \label{tab:vbxovpresults}
 \resizebox{\textwidth}{!}{
 \begin{tabular}{|l|llll|llll|llll|}
 \hline
 \multirow{2}{*}{\textbf{System}} &
   \multicolumn{4}{c|}{\textbf{AMI Eval}} &
   \multicolumn{4}{c|}{\textbf{Voxconverse Eval}} &
   \multicolumn{4}{c|}{\textbf{DISPLACE Eval}} \\ \cline{2-13} 
  &
   \textbf{FA} &
   \textbf{Miss} &
   \multicolumn{1}{l|}{\textbf{Conf.}} &
   \textbf{DER} &
   \textbf{FA} &
   \textbf{Miss} &
   \multicolumn{1}{l|}{\textbf{Conf.}} &
   \textbf{DER} &
   \textbf{FA} &
   \textbf{Miss} &
   \multicolumn{1}{l|}{\textbf{Conf.}} &
   \textbf{DER} \\ \hline
 AHC+VBx     & 0.0 & 15.6 &  \multicolumn{1}{l|}{11.9} & 27.43 & 0.0 & 3.1 & \multicolumn{1}{l|}{8.6} & 11.72 & 3.1 & 22.4 & \multicolumn{1}{l|}{11.15} & 36.75      \\ 
 AHC+Overlap+VBx    & 0.7  & 11.6  &  \multicolumn{1}{l|}{11.63} & 23.93 & 0.8 & 1.6 & \multicolumn{1}{l|}{8.45} & 10.85 & 3.8 & 21.0 & \multicolumn{1}{l|}{12.26} & 37.06       \\ \hline
 SC+VBx    & 0.0  & 15.6 & \multicolumn{1}{l|}{5.3 } & 20.90 & 0.0 & 3.1 & \multicolumn{1}{l|}{7.6} & 10.71 & 3.1 & 22.4 & \multicolumn{1}{l|}{10.38} &  35.98     \\ 
 SC+Overlap+VBx      & 0.7  & 11.6 &  \multicolumn{1}{l|}{6.7} & 19.00 & 0.8 & 1.6 & \multicolumn{1}{l|}{8.16} & 10.56 & 3.8 & 21.0 & \multicolumn{1}{l|}{11.12} &  35.92     \\ \hline
 SHARC + VBx   & 0.0 & 15.6 & \multicolumn{1}{l|}{4.8} & 20.34 & 0.0 & 3.1 & \multicolumn{1}{l|}{7.2} & 10.30  & 3.1 & 22.4 & \multicolumn{1}{l|}{6.29} &   31.89   \\ 
  \textcolor{black}{+ SHARC-Overlap + VBx}   & 0.7  & 11.6 & \multicolumn{1}{l|}{6.26 } & 18.56 & 0.8 & 1.6 & \multicolumn{1}{l|}{7.8} & 10.19 & 3.8 & 21.0 & \multicolumn{1}{l|}{ 6.77} & 31.57    \\ \hline
E-SHARC+VBx & 0.0 & 15.6 & \multicolumn{1}{l|}{3.68} & 19.16 & 0.0 & 3.1 & \multicolumn{1}{l|}{7.05} & 10.15 & 3.1 & 22.4 & \multicolumn{1}{l|}{ 6.8} & 32.61       \\ 
 \textcolor{black}{+ E-SHARC-Overlap + VBx} & {0.7} & {11.6} & \multicolumn{1}{l|}{{4.91}} & \textbf{17.21} & {0.8} & {1.6} & \multicolumn{1}{l|}{{7.7}} & \textbf{10.11} & 3.8 & {21.0} & \multicolumn{1}{l|}{ {6.6}} &  \textbf{31.40}     \\ \hline
 \end{tabular}}
 \end{table*}
\begin{table}[t!]
		\caption{DER (\%, with overlap $+$ without collar), DER* (\%, without overlap $+$ with 0.25s collar)  and DER** (\%, with overlap $+$ with 0.25s collar) comparison with state-of-the-art on AMI SDM , Voxconverse, and DISPLACE 2023 (phase 2) Evaluation datasets. prop.: proposed
}
		\label{tab:sota}
		\centering
		
		\resizebox{\columnwidth}{!}{
		\begin{tabular}{|l|c|c|c|c|} 
				\hline
				
				\multirow{1}{*}{\textbf{AMI SDM System}} 
			         &VAD & DER  & DER* & DER**\\ 
				\hline
				
				x-vec+AHC+VBx \cite{landini2020bayesian}  &Oracle  &27.4  & 12.6 & -\\
				SelfSup-PLDA-PIC +VBx \cite{singh2021self} &Oracle & 23.8  & 5.5 & -\\
                GAE-based+ SC \cite{WEI2023102991} &Oracle & - & 5.5 & - \\
                                  GADEC-based \cite{WEI2023102991} &Oracle & - & 4.2 & - \\
				E-SHARC (prop.)  &Oracle & 19.8 & 2.9 & -\\ 
				E-SHARC-Ovp +VBx (prop.) &Oracle & \textbf{17.2}  & \textbf{2.6} & -\\ 
                \cline{2-5}
                Pyannote \textcolor{black}{2.1.1} \cite{bredin21_interspeech}   &System & \textcolor{black}{27.1} & - & - \\
				Raj et al. \cite{raj2022gpu} &System & 23.7  & - & - \\
				Plaquet et al. \cite{plaquet23_interspeech} &System & \textbf{22.9} & - & -\\ 
                \textcolor{black}{ResNet x-vec.+VBx+OSD \cite{landini2024diaper}} &\textcolor{black}{System}  & \textcolor{black}{34.6} & - & - \\
				\textcolor{black}{DiaPer EEND \cite{landini2024diaper}} &\textcolor{black}{System}  &\textcolor{black}{37.5} & - & - \\
                   \textcolor{black}{E-SHARC (prop.)}  & \textcolor{black}{System}   & \textcolor{black}{28.2}  & \textcolor{black}{10.7} & - \\
				\textcolor{black}{+ E-SHARC-Ovp +VBx (prop.)}  & \textcolor{black}{System}   & \textcolor{black}{25.7} & \textcolor{black}{10.5} & -\\
												\hline
			       \multirow{1}{*}{\textbf{Voxconverse System}} 
				& &   &   & \\ \hline
				
				GAE-based+ SC \cite{WEI2023102991} &Oracle & - &8.0 & -\\
                GADEC-based \cite{WEI2023102991} &Oracle  & - & 7.7 & -\\
				E-SHARC (prop.)  &Oracle & 11.7 & 7.6 &9.2\\
                + E-SHARC-Ovp +VBx (prop.)  &Oracle &$\boldsymbol{10.1}$   & \textbf{6.3} & 7.5\\ 
                \cline{2-5}
                Pyannote \textcolor{black}{2.1.1} \cite{bredin21_interspeech}  &System & \textcolor{black}{11.2} & -  & - \\
				Plaquet et al. \cite{plaquet23_interspeech}  &System & \textbf{10.4}  & - & 5.8 \\
                \textcolor{black}{Pyannote VoxSRC2023 \cite{baroudi2023pyannote}} &\textcolor{black}{System} & -  & - & \textcolor{black}{\textbf{4.0}} \\
                \textcolor{black}{Krisp  VoxSRC2023 \cite{karamyan2023krisp}} &\textcolor{black}{System} & -  & - & \textcolor{black}{4.4} \\
                \textcolor{black}{ResNet x-vec.+VBx+OSD \cite{landini2024diaper}} &\textcolor{black}{System}  & - & - & \textcolor{black}{6.1} \\
                \textcolor{black}{DiaPer EEND \cite{landini2024diaper}} &\textcolor{black}{System}  & - & - & \textcolor{black}{18.2} \\
                \textcolor{black}{E-SHARC (prop.)}  & \textcolor{black}{System}   & \textcolor{black}{16.1} &\textcolor{black}{9.7}  & \textcolor{black}{11.3} \\
                \textcolor{black}{+ E-SHARC-Ovp +VBx (prop.)}  &\textcolor{black}{System} & \textcolor{black}{15.1}  &\textcolor{black}{8.8}  & \textcolor{black}{10.0} \\
				\hline
				 \multirow{1}{*}{\textbf{DISPLACE System}} 
				& &     &  & \\ \hline
				Baseline \cite{baghel2023displace,baghel2023summary}   & Baseline  & 32.2   & - & -\\ 
				E-SHARC-Ovp +VBx (prop.)  &Baseline & 31.4 & - & - \\	
             E-SHARC-Ovp +VBx (prop.)  & System  & 29.9  & - & -\\
				Winning system \cite{baghel2023summary} & System &  \textbf{27.8}   & - & - \\ 
				  \hline
		\end{tabular}}
	\end{table}
\subsection{Implementation details}
\subsubsection {{Optimizer}}
The SHARC model\footnote{https://github.com/prachiisc/SHARC.git} is trained with a Stochastic Gradient Descent (SGD) optimizer with a learning rate $lr$=$0.01$ (for Voxconverse) and $lr$=$1e$-$3$ (for AMI) for $500$ epochs. Similarly, the E-SHARC is also trained with an SGD optimizer. In this case, the learning rate is $1e$-$6$ for the ETDNN module and $1e$-$3$ for the GNN module; the model is trained for 20 epochs. SHARC-Overlap is initialized with SHARC model weights and trained with $lr$=$1e$-$3$ for 100 epochs. E-SHARC-Overlap is initialized with E-SHARC model weights and trained for $5$ additional epochs.

\subsubsection {{Adjacency matrix}}
The graph adjacency matrix of each recording is obtained using a similarity matrix with k highest similarities in each row. At each level of the hierarchy, the x-vectors, also called the node identity features, are passed to the PLDA model followed by sigmoid activation to generate the similarity score matrix. 

\section{Results}
The performance of the system is measured based on Diarization Error Rate (DER) \cite{fiscus2006rich}. It is the sum of false alarm, miss rate, and speaker confusion error. The implementation is based on \emph{dscore toolkit}\footnote{https://github.com/nryant/dscore}. We have used ground truth speech activity decisions for evaluating AMI and Voxconverse datasets. \textcolor{black}{However, we have integrated voice activity detection (VAD) system called Pyannote VAD \cite{bredin21_interspeech} in our final model to compare with the other state-of-the-art approaches.}

We also evaluate our trained model on the DISPLACE challenge dataset. As per challenge guidelines, we have used the DISPLACE challenge baseline VAD model. The model is based on TDNN architecture \cite{baghel2023displace,ryant21_interspeech} with speech and non-speech classification. The Voxconverse E-SHARC model is also used for DISPLACE evaluations as the DISPLACE dataset does not contain a training partition.

\subsection{Comparison with  baseline systems}
The best values of hyper-parameters $k$ and $\tau$ are obtained based on DER on the dev set, applied on the eval set, and vice versa for all the datasets. Table \ref{tab:hyperparameters} shows the values of hyperparameters used for the AMI and Voxconverse experiments.

Table \ref{tab:purityCov} reports the performance based on the cluster purity and coverage on Voxconverse dev set using pyannote-metric~\cite{pyannote.metrics}. \textbf{Cluster purity} is defined as the percentage of segments from predicted speakers belonging to the corresponding speaker in the ground truth. \textbf{Cluster coverage} is defined as the percentage of segments from the ground truth speaker covered by the predicted speaker. 
From the table, it can be observed that the baseline with AHC has high purity but low coverage. However, the baseline with spectral clustering (SC) has lower purity but higher coverage. \textcolor{black}{In our proposed approach, both the purity and the coverage are high, indicating that supervised clustering can achieve better trade-off}.

Table \ref{tab:ovpresults} shows the false alarm (FA), miss, and confusion error along with the overall DER metric of baseline AHC, SC approaches, and the proposed SHARC/E-SHARC models with and without overlap assignment for three different datasets: AMI Single Distant Microphone (SDM), Voxconverse and DISPLACE challenge datasets. From Table \ref{tab:ovpresults}, it can be observed that SHARC and E-SHARC perform significantly better than AHC and SC baselines on all the datasets.  \textcolor{black}{E-SHARC achieves $11\%$, $13\%$ and $20\%$ relative improvements over the best baseline on AMI, Voxconverse and DISPLACE datasets, respectively}. 

To fairly compare the performance of SHARC-Overlap and E-SHARC-Overlap with the baseline clustering approaches, we have also integrated the overlap assignment for AHC and SC, which is similar to the E-SHARC-Overlap approach. \\
\textbf{Baseline overlap assignment (AHC/SC+Overlap):} The first pass clustering (AHC/SC) is performed using the PLDA similarity scores matrix to generate the parent speaker labels for each x-vector. The pyannote overlap detector is used to select the x-vector segments containing overlaps. The within-cluster similarity scores are set to the lowest value for these overlapping segments in the matrix. Then, the top $k'$ similarity scores of each overlapping x-vector are selected for the second pass clustering. The parent speaker labels are assigned for these top $k'$ scores. The mode of the $k'$ parent speaker labels is the second speaker assigned to the overlapping regions. The value of $k'$  is selected as 30 based on validation experiments for SHARC-Overlap and E-SHARC-Overlap. 

\begin{figure*}[t!]
  \centering
  \includegraphics[trim={6cm 0cm 9.5cm 0cm},clip,width=0.75\textwidth]{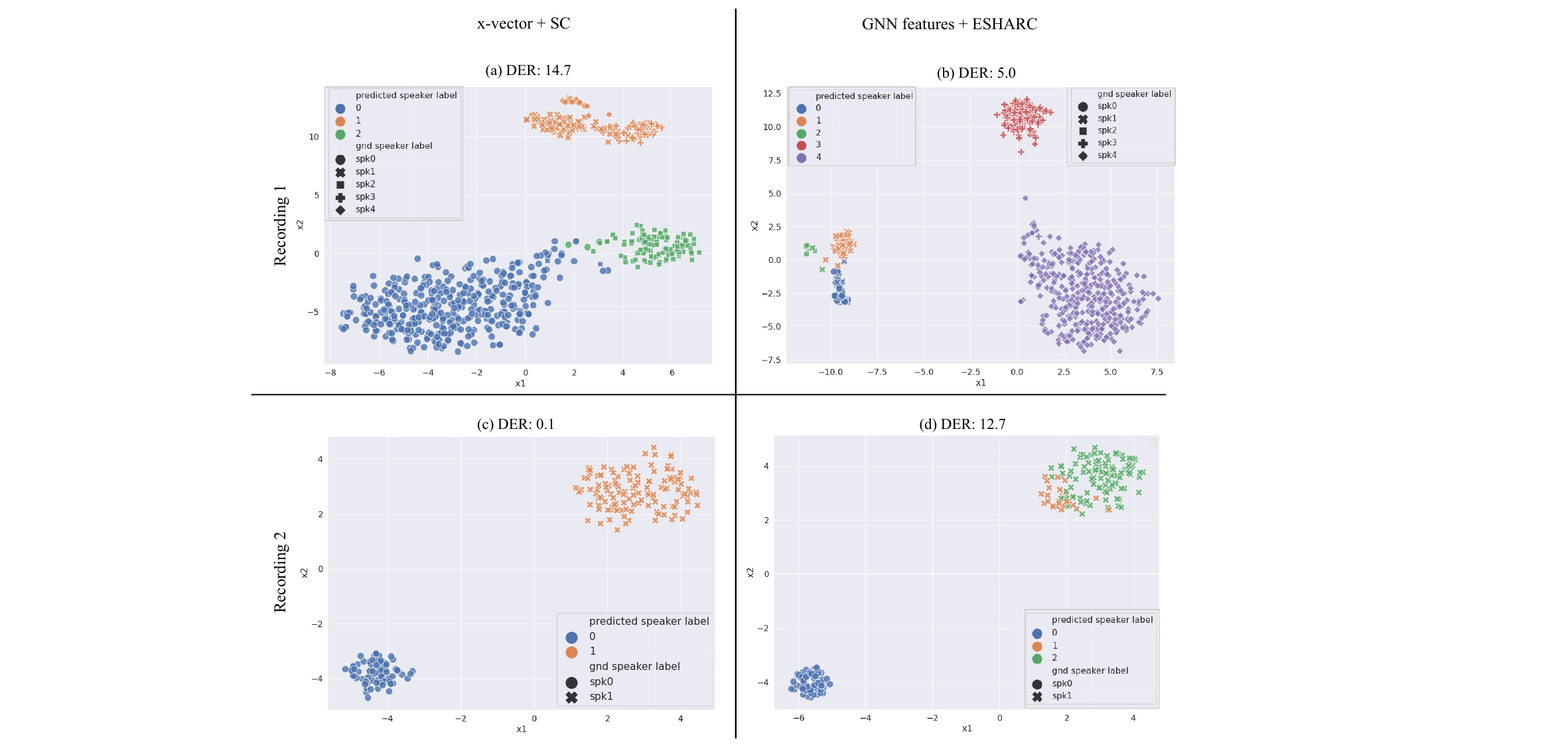}
  \vspace{-10pt}
  \caption{2D t-SNE plot to compare x-vectors and GNN embeddings for two different recordings from Voxconverse dev set.   Ground truth labels are represented as shapes while predictions are represented as colors. In both cases, the proposed E-SHARC yields representations with improved separability. However, the DER for recording-2 deteriorated due to early stopping. }
  \label{fig:tsne1}
    \vspace{-10pt}
\end{figure*}

\textcolor{black}{From Table \ref{tab:ovpresults}, it can be observed that adding overlap assignment improves the DER performance for all models.}
\textcolor{black} {We achieve $11.6\%$, 
$5.2\%$ , and $19.8\%$ relative improvements for AMI, Voxconverse, and DISPLACE, respectively, over the best baseline system. }

Table \ref{tab:vbxovpresults} shows the results of baseline and proposed approaches with VBx resegmentation \cite{landini2020bayesian}. \textcolor{black}{The VBx is initialized with the corresponding clustering output of the SHARC/ESHARC models.} This is followed by a re-segmentation step using the HMM-PLDA model. It can be observed that VBx segmentation improves performance as it helps refine the speaker boundaries based on temporal information. 
 After the VBx iterations, the second speaker is predicted for the overlapping regions based on the second-highest posterior probability. The E-SHARC-Overlap + VBx achieves the lowest DER for all the datasets, resulting in  $9.4\%$, $4.3\%$, and $12.6\%$ relative improvements for AMI, Voxconverse, and DISPLACE, respectively over the  best baseline. 
\begin{figure}[t!]
  \centering
  \includegraphics[trim={0cm 0cm 0cm 0cm},clip,width=\columnwidth]{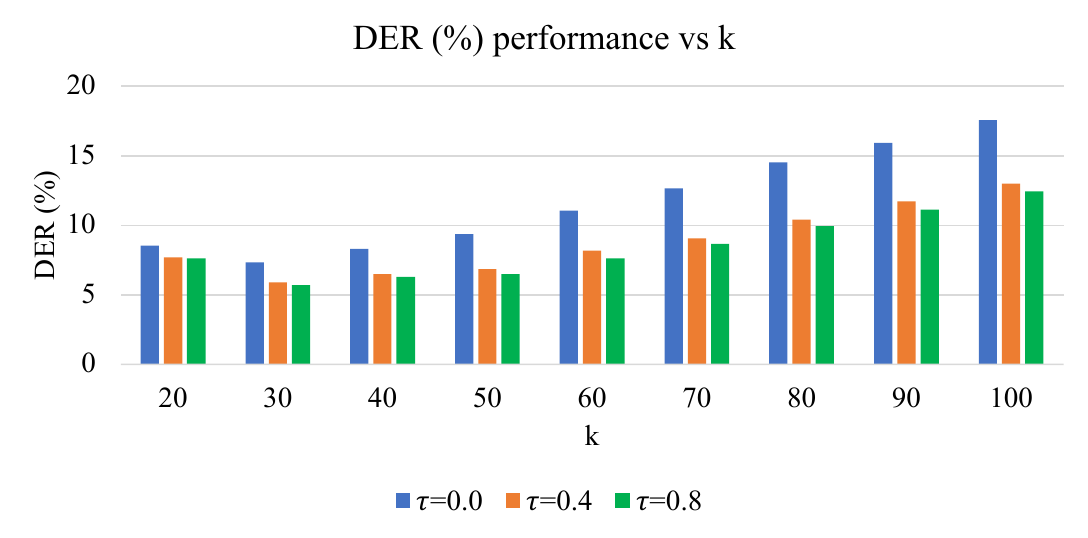}
  \vspace{-20pt}
  \caption{Plot comparing DER performance for k ranging from 20-100 and $\tau \in \{0.0, 0.4, 0.8\}$ for Voxconverse dev set. }
  \label{fig:2dplotfilewise}
    \vspace{-20pt}
\end{figure}

\subsection{Comparison with other published works}
Table \ref{tab:sota} compares the proposed approach with state-of-the-art approaches for AMI, Voxconverse, and DISPLACE datasets using different VAD configurations. Pyannote \cite{bredin21_interspeech} is the end-to-end pyannote model with overlap detection modules. 
 Raj et al. \cite{raj2022gpu}, and Plaquet et al. \cite{plaquet23_interspeech} explore the recent end-to-end models for diarization. The work reported in \cite{WEI2023102991} proposed Graph Attention-Based Deep Embedded Clustering (GADEC), which performs graph attention-based clustering using multi-objective training.  It also shows the results of the Graph attentional encoder (GAE) based approach for metric learning followed by spectral clustering.  {Our E-SHARC-Ovp + VBx outperforms the state-of-the-art results for AMI SDM and Voxconverse systems using oracle VAD, although this setting is somewhat deprecated in the recent evaluations.} \textcolor{black}{We have also reported performance of systems of VoxSRC 2023 challenge \cite{baroudi2023pyannote,karamyan2023krisp} for Voxconverse datasets, which involves more complex models and larger training data for diarization. We have reported results of ResNet x-vector models  with system VAD, VBx and OSD \cite{landini2024diaper}. We have also compared with recent \textcolor{black}{Perceiver based end-to-end diarization system (DiaPer)} \cite{landini2024diaper}, trained with $2500$ hours of simulated data as opposed to $75$/$400$ hours of real/simulated data for AMI/Voxconverse in the proposed approach.}  In the case of DISPLACE, the winning system investigated different VAD and model combination strategies, while the proposed system was trained on out-of-set (Voxceleb) data. {\color{black} The results suggest that the E-SHARC approach performs better than the DiaPER EEND system and x-vec + VBx system in case of AMI SDM system.
This may be attributed to the improved modeling of the intra and inter speaker relationships through the graph clustering. 
However, Plaquet et al. achieves the best results in this setting. In that work, the authors use a combination of EEND and clustering jointly. In such a case, the EEND performs efficient modeling within a small window with overlap handling and VAD,  while a linking/clustering across windows is performed by a spectral clustering setting.   As a future scope, it is worthwhile to explore the incorporation  of EEND with our proposal of graph clustering. We hypothesize that such a framework could further improve the results and combine the advantages of graphs with EEND setting.}
 
 


\begin{figure*}[t!]
  \centering
  \includegraphics[trim={0cm 6cm 7cm 5cm},clip,width=\textwidth]{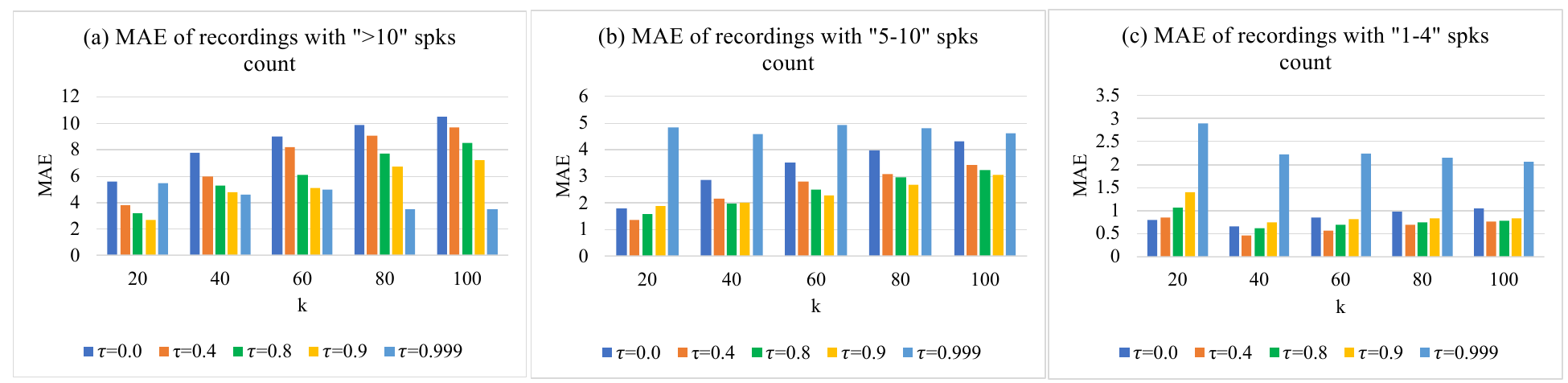}
  \vspace{-10pt}
  \caption{{\color{black}Comparison of mean absolute error (MAE) for speaker counting task for different values of  $k$  and $\tau$ on Voxconverse Dev. data }}
    \vspace{-10pt}
  \label{fig:plotMSEfilewise}
\end{figure*}



\section{Ablation Studies}
\subsection{Choice of hyper-parameters}
\label{hyperparameters}

 \textit{SHARC parameters ($k$, $\tau$):}  A high value of $k$ while training allows to capture more edges in the graph and improves the model predictions. The parameter
$\tau$ (edge prediction probability threshold) sets the minimum required probability to allow the edge connection between nodes. 
Figure \ref{fig:2dplotfilewise} shows the impact of different values of $k$ and $\tau$ on DER on the Voxconverse dev set. As $k$ increases, the DER increases as higher $k$ generates fewer speakers. On the other hand, the DER is higher for a smaller value of $\tau$ ($\tau < 0.2$). The best values of $k$ and $\tau$ are $30$ and $0.8$. 
\subsection{Speaker counting task}
We also perform the speaker counting task evaluation using the proposed model. 
\textcolor{black}{Figure \ref{fig:plotMSEfilewise}} shows the performance of the speaker counting task for the Voxconverse dataset. It shows the impact of $k$ and $\tau$ on the mean absolute error (MAE) between the ground truth number of speakers and the predicted number of speakers. 
The figure is divided into three categories of recordings based on the number of speakers present. Plot (a) shows the MAE for recordings with a higher number of speakers ($>10$). As $k$ increases, MAE increases, and as $\tau$ increases, MAE reduces. Plot (b) and (c) shows the MAE for recordings containing a small number of speakers where the trends are different. In these settings, MAE is relatively low across different $k$ and $\tau$. \textcolor{black}{A higher value of $k$ while testing results in fewer levels and clusters, while a higher value of $\tau$ leads to fewer connections and generates a higher number of clusters. Therefore, for plots (a) and (b), lower  $k$ and higher $\tau$ leads to better MAE, but for plot (c) the trend is different where  the best values of $k$ and $\tau$ are relatively higher and lower, respectively.}


\begin{figure*}[t!]
  \centering
  \includegraphics[trim={0cm 4cm 0cm 5cm},clip,width=\textwidth]{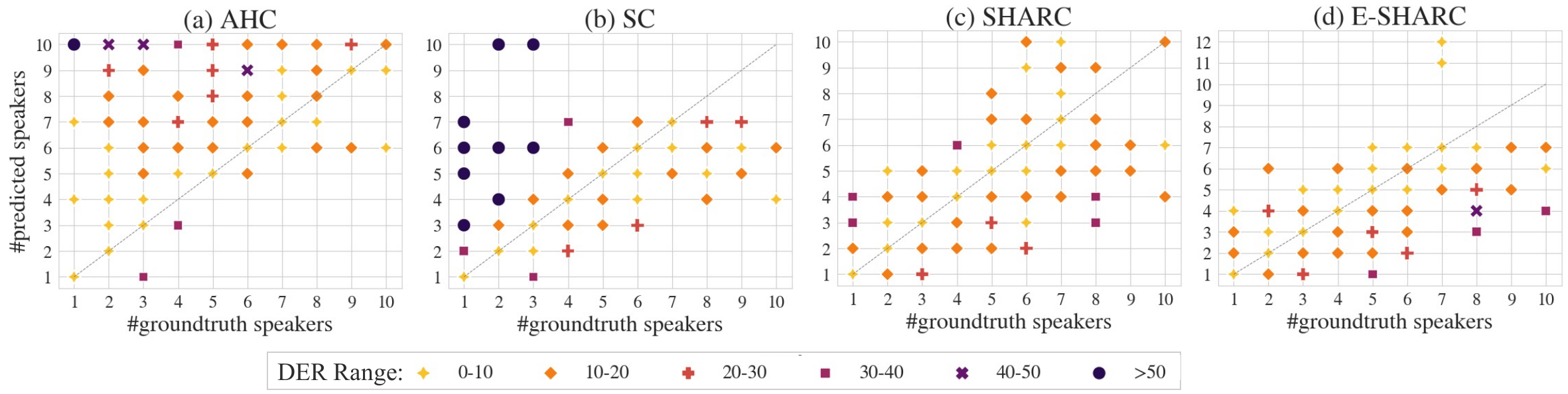}
  \caption{\textcolor{black}{2D scatter plot showing \# of ground truth speakers vs \# of predicted speakers using (a) AHC, (b) SC, (c) SHARC  and (d) E-SHARC models on Voxconverse Dev. data. The different colors represent different ranges of average DER (\%).}}
  \label{fig:plotscatter}
\end{figure*}

\textcolor{black}{\subsection{GNN vs LSTM architectures}
Similar to GNNs, other neural architectures like LSTM may also be capable of performing speaker embedding extraction and clustering \cite{wang2018speaker}. 
We have  experimented with  bidirectional LSTM (BLSTM) architecture, where all the other data and modeling settings are kept identical to the proposal of SHARC. We replaced the GNN layer with BLSTM layer in the SHARC model such that number of parameters are similar. The results, shown in Table \ref{tab:GNNLSTM}, indicate that the GNN based SHARC performs better than the BLSTM architecture for the AMI diarization task. This experiment illustrates the improved representation learning achieved using the GNN for speaker diarization.}

\begin{table}[t!]
\begin{center}
\textcolor{black}{\caption{Performance comparison of GNN vs BLSTM architecture in SHARC on AMI SDM Eval set using oracle VAD. }}
 \label{tab:GNNLSTM}
\begin{tabular}{l|l|l|l}
\hline 
\textcolor{black}{\textbf{Architecture}} & \textcolor{black}{\textbf{\# Hid. Dim}} & \textcolor{black}{\textbf{\# Parameters} }& \textcolor{black}{\textbf{DER \%}} \\ \hline
\textcolor{black}{  GNN}           & \textcolor{black}{2048}           & \textcolor{black}{8M}           & \textcolor{black}{\textbf{21.27}}        \\
\textcolor{black}{ BLSTM}        & \textcolor{black}{2048}           & \textcolor{black}{16M}          & \textcolor{black}{22.92}        \\
\textcolor{black}{ BLSTM}         & \textcolor{black}{1024}           & \textcolor{black}{8M}           & \textcolor{black}{22.13}    \\   
\hline 
\end{tabular}
\end{center}
\vspace {-18pt}
\end{table}
\subsection{Representation visualization}

Figure \ref{fig:tsne1} shows the t-SNE plots for two recordings (two rows) from Voxconverse dev set for baseline and the proposed E-SHARC approach. The first column shows a t-SNE plot of x-vectors with SC, and the second column shows GNN features with E-SHARC. Recording 1 contains five speakers represented by different shapes. The SC predicts only three speakers out of five, as shown in different colors. However, E-SHARC can form five different clusters and provide a lower DER. It can also be observed that the within-cluster covariance is lower for the GNN features compared to x-vectors. For recording 2, the SC can predict the correct number of speakers. On the other hand, E-SHARC splits spk1 from ground truth into two different speakers shown in different colors. Although the features are well separated in terms of speaker clusters, the stopping criterion has resulted in early stopping.   

\textcolor{black}{Figure \ref{fig:plotscatter} shows a 2D scatter plot comparing the performance of AHC, SC, SHARC and E-SHARC in terms of the predicted v/s true number of speakers on Voxconverse dataset. 
It can be observed that, the AHC and the SC predict more number of speakers compared to ground-truth resulting in  high DERs ($> 50$\%).
On the other side, the SHARC and the E-SHARC models 
are more accurate in predicting the  true number of speakers, resulting in  a lower file-wise DERs.}


\section{Conclusion}
\textcolor{black}{This paper extends our prior work \cite{singh2023supervised} and provides a detailed analysis of  end-to-end supervised hierarchical neural graph clustering approach for speaker diarization.} The proposed approach involves supervised graph model training with a clustering-based loss. This model, called E-SHARC, performs hierarchical clustering using a GNN model by minimizing intra-speaker distances and maximizing inter-speaker distances. The model is further extended to perform overlapped speaker prediction.
This enables the prediction and assignment of multiple speakers in the overlapped speech regions, further improving the diarization performance. We also showed the effectiveness of the proposed approach on a challenging code-mixed DISPLACE dataset.

\section*{Acknowledgments}
The authors would like to thank Amrit Kaul for helping in the initial experiments. The authors would like to thank Michael Free, Rohit Singh, and Shakti Srivastava from BT Research for their valuable input. The authors also thank Herv\'e Bredin for providing pretrained pyannote VAD and Overlap detection models.

\bibliographystyle{IEEEtran}
\bibliography{TASLP_jrnl}


\newpage

%
%
%
%

\vfill

\end{document}